\documentclass[twocolumn,showpacs,amsmath,amssymb,aps,prb,superscriptaddress]{revtex4}
\usepackage{graphicx}     
\usepackage{bm}

\begin{document}

\title{Topological order in paired states of fermions in two-dimensions with breaking of parity and time-reversal symmetries}

\author{Noah Bray-Ali}
\affiliation{Department of Physics and Astronomy,
University of Kentucky, Lexington, KY 40506}

\author{Letian Ding}
\affiliation{Department of Physics and Astronomy, University of
Southern California, Los Angeles, CA 90089}

\author{Stephan Haas}
\affiliation{Department of Physics and Astronomy, University of
Southern California, Los Angeles, CA 90089}

\date{\today}

\begin{abstract}
We numerically evaluate the entanglement spectrum (singular value decomposition of the wavefunction) of paired states of fermions in two dimensions that break parity and time-reversal symmetries, focusing on the spin-polarized $p_x+ip_y$ case.  The entanglement spectrum of the weak-pairing (BCS) phase contains a Majorana zero mode, indicating non-Abelian topological order.  In contrast, for the strong-pairing (BEC) phase, we find no such mode, consistent with Abelian topological order.  
\end{abstract}


\maketitle

\section{Introduction}
Two-dimensional fermion systems with pairing that breaks parity and time-reversal symmetries come in a variety of forms including quantum hall fluids,\cite{willett87} superfluids,\cite{osheroff} superconductors,\cite{sigrist} and condensates of cold atoms near a Feshbach resonance.\cite{gurarie}  For spin-polarized fermions, the simplest pairing order parameter that breaks these symmetries, $\Delta_{\bf p} \propto p_x+ip_y,$ depends on the relative momentum ${\bf p}$ of the fermions in a pair.  For momentum independent, $s$-wave pairing, a smooth cross-over occurs from weak-pairing (BCS) to strong-pairing (BEC).  In the $p_x+ip_y$ case, the two phases have different topological order and are separated by a quantum phase transition.\cite{read99}  

Recent proposals for fault-tolerant quantum computation and information processing rely on topological order in fermion systems with $p_x+ip_y$ pairing,\cite{tewari06} but detecting and characterizing such order remain open problems.  For example, the symmetry and bulk spectral properties of the BCS and BEC phases are identical, but they have dramatically different topological order: quantum vortices have non-Abelian statistics in the weak-pairing phase and Abelian statistics in the strong-pairing phase.  We apply ideas from quantum information to investigate topological order in these interesting, paired fermion systems.  

The entanglement spectrum\cite{haldane08} and the entanglement entropy\cite{chuang} contain information about the universal properties of a quantum state.  We define them by dividing the system into a block $A$ with feature size $L$ and an environment $B$, and then performing a Schmidt decomposition,
\begin{equation}
|\psi\rangle=\sum_i e^{-\frac{1}{2} \xi_i} |\psi_i^A\rangle\otimes|\psi_i^B \rangle.
\label{schmidt}
\end{equation}
Here, the orthonormal sets of states $\{|\psi_i^A\rangle\}$, $\{|\psi_i^B\rangle\}$ span $A$ and $B.$  The entanglement spectrum  $\{\xi_i\}$ gives the entanglement entropy $S=\sum_i \xi_i e^{-\xi_i}$.


In this Letter, we report the first large-scale numerical calculations of the entanglement entropy and spectrum of two-dimensional fermion systems with $p_x+ip_y$ pairing.  We find that the entanglement spectrum qualitatively distinguishes the topological order occurring in the two phases.  In particular, we find that the low-lying spectrum in the weak-pairing phase contains a chiral, gapless fermion excitation.  The weak-pairing phase is known to have a chiral, gapless Majorana edge mode.\cite{read99}  This mode is related to the Majorana zero mode that appears in vortex cores and gives vortices non-Abelian statistics.\cite{read99,ivanov01}

 We reduce the problem of evaluating the entanglement spectrum and entanglement entropy to diagonalizing a quadratic entanglement Hamiltonian.\cite{chung01}  This approach does not include fluctuations of the pairing order parameter, and, hence, we do not expect to observe a universal, topological term in the entanglement entropy\cite{kitaev06} in either the weak-pairing or strong-pairing phase,\cite{nussinov09} despite the fact that both phases have non-trivial quantum dimension $D=2$.  Indeed, we confirm that the size of the leading correction term depends on the geometry of the block, and is in fact proportional to the number of corners.\cite{kumar07}  In contrast, the entanglement spectrum detects non-Abelian topological order in the ground-state wavefunction for states of paired fermions even when pairing fluctuations are neglected.

\section{Pairing Hamiltonian}  The following BCS Hamiltonian\cite{bcs57} serves as a minimal model for a single band of spin-polarized fermions with $p_x+ip_y$ pairing on a square lattice:
\begin{eqnarray}
H=\sum_{\langle {\bm r},{\bm r'}\rangle}  \left(-t c^{\dagger}_{{\bm r}}c_{{\bm r'}} - \gamma_{{\bm r},{\bm r'}}
c^{\dagger}_{{\bm r}}c^{\dagger}_{{\bm r'}} + h.c. \right) +2\lambda \sum_{{\bm r}}
c^{\dagger}_{{\bm r}}c_{{\bm r}}
\label{rspace}
\end{eqnarray}
We consider only nearest-neighbor $\langle {\bm r}, {\bm r'}\rangle$ hopping $t$ and pairing $\gamma_{{\bm r},{\bm r'}}$ interactions.  The hopping strength $t$ and coupling $\lambda$ are taken to be real and positive, without loss of generality.  The pairing interaction $\gamma_{{\bm r},{\bm r'}}$ breaks both time-reversal and parity symmetries: $\gamma_{{\bm r},{\bm r}+\hat{x}}=-\gamma_{{\bm r},{\bm r}-\hat{x}}=i\gamma_{{\bm r},{\bm r}+\hat{y}}=-i\gamma_{{\bm r},{\bm r}-\hat{y}}=i\gamma	$.   Here, $\gamma$ is real and $\hat{x},\hat{y}$ are the primitive translation vectors of the square lattice.  We use periodic boundary conditions in our numerical calculations.

\begin{figure}
\includegraphics[width=3.4in]{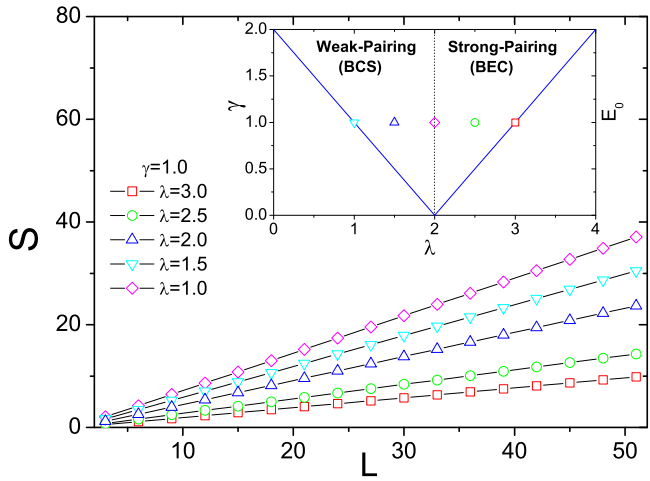}
\caption{(Color online) Entanglement entropy $S$ between a square of side length $L$ and its environment as a function of $\lambda$ at fixed pairing strength $\gamma=1.0$. (Inset) The zero-temperature phase diagram of two-dimensional fermions with $p_x+ip_y$ pairing and plot of the bulk spectral gap $E_0$.  The phase boundary between weak-pairing and strong-pairing is the vertical $\gamma$-independent line at $\lambda_c=2t$. The spectral gap vanishes at the critical coupling and grows linearly with $|\lambda-\lambda_c|$.  Data points indicate the parameters chosen in our numerical calculations (t=1).}
	\label{area}
\end{figure}

The pairing Hamiltonian (\ref{rspace}) is quadratic and can be solved exactly using a Bogoliubov transformation,\cite{blaizot} yielding the phase diagram shown in the inset of Fig.~\ref{area}.\cite{read99}  The critical line at $\lambda_c=2t$, separates the weak-pairing (BCS) phase from the strong-pairing (BEC) phase.  Both phases have a spectral gap $E_0=t|\lambda-\lambda_c|$ to bulk excitations show in the inset of Fig.~\ref{area} and  determined by minimizing the Bogoliubov quasi-particle dispersion:$E_{\bm p}= \sqrt{\xi_{\bm p} ^2 + \left | \Delta_{\bm p} \right |^2}.$  The pairing order parameter $\Delta_{\bm p} =2\gamma(\sin p_{x}+ i\sin p_{y})$ transforms under the symmetries of the square lattice in the same way as an $\ell=1, \ell^z=1$ spherical harmonic.   At small $p$, we expand $\Delta_{\bm p}\propto p_x+ ip_y$, and see the $p_x+ip_y$ pairing explicitly.  Similarly, at small $p$, the single-particle kinetic energy  $\xi_{\bm p} = -2t(\cos p_x + \cos p_y)+2\lambda$, takes the form $\xi_{\bm p}=p^2/2m^*- \mu$, with effective mass $m^*=1/2t$ and $\mu=4t-2\lambda$.  The weak-pairing phase $\lambda<\lambda_c$ corresponds to $\mu>0$, while strong-pairing $\lambda>\lambda_c$ corresponds to $\mu<0$.  Near the quantum phase transition $\mu=0$, the low-energy spectrum $E_{\bm p}=\sqrt{4\gamma^2 p^2 + \mu^2}$ has a relativistic form with $2\gamma$ playing the role of the speed of light.

\section{Entanglement Hamiltonian}
The two-point correlation functions provide a complete description of the ground state of the quadratic Hamiltonian (\ref{rspace}), and allow an efficient numerical evaluation of the Schmidt decomposition (\ref{schmidt}).\cite{chung01}  In fact, the Schmidt decomposition of the pairing Hamiltonian ground-state reduces to diagonalizing the following entanglement Hamiltonian $H_e$ which acts on the sites of the block $A$:\cite{blaizot}
\begin{eqnarray}
H_e=\sum_{{\bm r},{\bm r'}} C_{{\bm r},{\bm r'}} \left(c^{\dagger}_{{\bm r}}c_{{\bm r'}} + h.c. \right)
+\sum_{{\bm r},{\bm r'}}\left (F_{{\bm r},{\bm r'}}
c^{\dagger}_{{\bm r}}c^{\dagger}_{{\bm r'}}+h.c. \right).
\label{he}
\end{eqnarray}
Here, in contrast to (\ref{rspace}), the hopping parameters $C_{{\bm r},{\bm r'}}=\int d^2p/(2\pi)^2   e^{i{\bm p} \cdot({\bm r}-{\bm r'})} (E_{\bm p} -\xi_{\bm p})/2 E_{\bm p}$ and pairing parameters $F_{{\bm r},{\bm r'}}=\int d^2p/(2\pi)^2 e^{i{\bm p}\cdot ({\bm r}-{\bm r'})} \Delta_{\bm p}/2E_{\bm p}$ extend beyond nearest-neighbors and are given by the two-point correlation functions in the ground state of the pairing Hamiltonian (\ref{rspace}).  The entanglement Hamiltonian is  quadratic, and can be exactly solved by numerically performing a Bogoliubov transformation to the quasi-particle operators $\alpha_n$, for $n=\pm 1,\pm 2,\ldots \pm N_A$, where, $N_A$ is the number of sites in the block $A$.\cite{blaizot}  In terms of the quasi-particles, the entanglement Hamiltonian has the form $H_e=\sum_{n>0} f(\epsilon_n)\alpha^{\dagger}_{n }\alpha_n$, where, $f(\epsilon)=(e^\epsilon+1)^{-1}$ is the Fermi function and the quasi-particle block energies $\{\epsilon_n\}$ generate the entanglement spectrum.  In particular, the entanglement entropy is given by $S=-\sum_n f(\epsilon_n)\log f(\epsilon_n)$.

\begin{figure}
\includegraphics[width=3.6in]{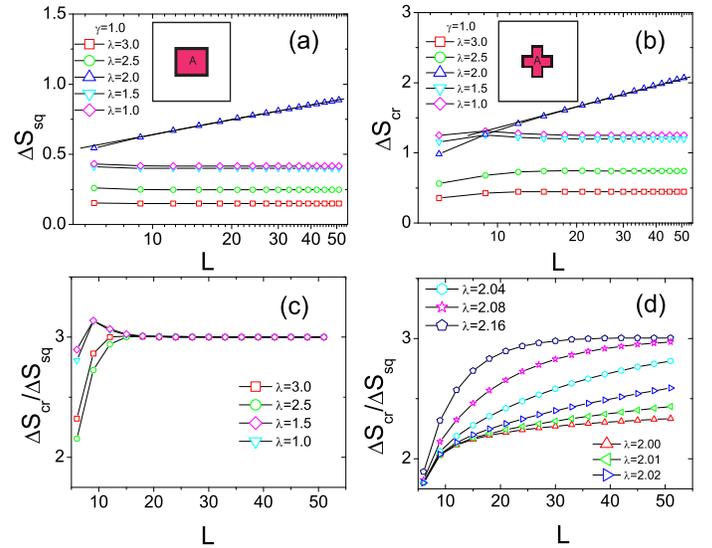} 
	\caption{(Color online) Leading correction term $\Delta S$ to the perimeter law for a square (a) and cross-shaped (b) partition as a function of block size $L$. (Inset) geometry of the partitions.  Notice the linear scale for $\Delta S$ and the logarithmic scale for $L$ in both (a) and (b).  Solid lines are guides to the eye.  Ratio $\Delta S_{cr}/\Delta S_{sq}$ of the leading correction terms from (a) and (b) as function of block size $L$: (c) within the weak-pairing and strong-pairing phases; (d) approaching the quantum phase transition from the strong-pairing regime. }
\label{sub}
\end{figure}


\section{Results} The entanglement entropy $S$ as a function of the block size $L$ is shown in Fig.~\ref{area}.  We consider various $\lambda$ sweeping through the quantum phase transition, as shown in the inset.  The entropy grows linearly with $L$ for this two-dimensional system.  We interpret this as a perimeter law $S_L = aL+\ldots$ , where, the ratio of the correction terms to $L$ vanishes in the limit $L\rightarrow\infty$.  
Our large-scale numerical results agree with general arguments that a perimeter law must hold in the gapped phases.\cite{hastings}  At the quantum critical point, the gap vanishes at a Majorana point,\cite{read99} and no theoretical predictions or previous numerical results are available.

Using these large-scale numerical results, we are able to extract the leading correction to the perimeter law $\Delta S=-3(S-aL)$.\cite{prl}  We plot the size dependence of the leading correction $\Delta S_{sq}$ for the square shaped partition shown in Fig.~\ref{sub}(a) and for the cross-shaped partition $\Delta S_{cr}$ shown in Fig.~\ref{sub}(b).  For both geometries, the leading correction grows at the critical point with $L$, without sign of saturation.  By contrast, in the weak-pairing and strong-pairing phases, the leading correction saturates to an $L$ independent value as $L\rightarrow\infty$.  We interpret the growth at the critical point as a logarithmic divergence, of the form $S=aL-b\log L +\ldots$ .  This is the first indication that a Majorana point exhibits a logarithmic correction to the perimeter law for the entanglement entropy, although additive logarithmic corrections have been observed in two-dimensional systems with other kinds of nodal excitations.\cite{moore06,prl} 

In Fig.~\ref{sub}(c) and (d), we analyze the geometry dependence by plotting the ratio of the leading correction $\Delta S_{cr}/\Delta S_{sq}$ for the two partition geometries.  In both strong-pairing and weak-pairing phases (Fig.~\ref{sub}c), the ratio $\Delta S_{cr}/\Delta S_{sq}\rightarrow 3$ approaches the ratio of the number of corners in the cross partition to the number in the square partition.  We have examined other geometries and find the behavior $\Delta S=cn_c$, where, $n_c$ is the number of corners and $c$ is a positive coefficient.\cite{kumar07}  In contrast, when pairing fluctuations are allowed, the topological term $\Delta S=3\log 2$ has no geometry dependence.\cite{kitaev06}  To check that our results reflect the asymptotic behavior of the system, we reduce the de-tuning from the critical point while staying on the BEC side (See Fig. 2(d)).   For system size $L\gg\xi$ much bigger than the diverging length scale $\xi= 2\gamma/|\mu|\propto|\lambda-\lambda_c|^{-1}$, the behavior $\Delta S_{cr}/\Delta S_{sq}\rightarrow 3$ observed deep within the gapped phases (Fig. 2(c)) emerges near the critical point as well.  Thus, these large-scale numerical simulations indicate a geometric origin of the leading correction to the perimeter law for the entanglement entropy.

\begin{figure}
\includegraphics[width=3.6in]{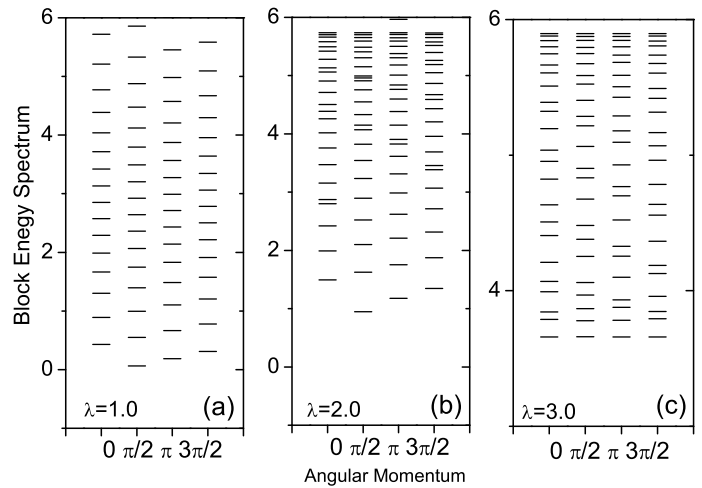}
\caption{Low-lying quasi-particle entanglement spectrum $\{\epsilon_n\}$ (a) in the weak-pairing phase, (b) at the quantum phase transition, and (c) in the strong-pairing phase with fixed pairing strength $\gamma=1.0$ and system size $L=24$.  We divide the spectrum into four sectors, corresponding to the irreducible representations of the point group of the square lattice and labeled by the phase factor acquired by the quasi-particle wavefunction during a $\pi/2$ rotation.}

\label{spectrum_finite}
\end{figure}

To detect topological order, we turn to the entanglement spectrum shown in Fig.~\ref{spectrum_finite}.  Now, in the weak-pairing phase, the energy spectrum of the pairing Hamiltonian (\ref{rspace}) for a system in the form of a disc of radius $R$ contains a chiral fermion edge mode with energy $E\propto m/R$ proportional to angular momentum $m$.\cite{read99}  To detect such a mode in the square geometry, one must label the quasi-particle block energies $\{\epsilon_n\}$ by the phase factor $\phi_n=0,\pi/2,\pi,3\pi/2$ acquired by the quasi-particle wavefunction under the elementary $\pi/2$ rotation symmetry of the square lattice.  This phase factor plays the role of angular momentum in a lattice system.  

In the weak-pairing phase, Fig.~\ref{spectrum_finite}(a) we find that both the energy $\epsilon_n\propto n$ and the phase factor $2 \phi_n/\pi=n ({\rm mod} 4)$ are proportional to the level index $n=1,2,\ldots$.  Eliminating the level index, we find $\epsilon_n\propto \phi_n$.  This is precisely the relationship expected for a gapless chiral mode, and observed in the weak-pairing phase of the pairing Hamiltonian (\ref{rspace}) along an edge.\cite{read99}  In the strong-pairing phase, Fig.~\ref{spectrum_finite}(c), the phase factor $\phi_n$ and level index have no apparent relationship.  At the critical point Fig.~\ref{spectrum_finite}(b) the phase factor and level index are proportional for the lowest levels, but have no relationship for higher quasi-particle block energy.   We contrast the dispersionless low-lying spectrum in the strong-pairing phase with the linearly dispersing spectrum in the weak-pairing phase, and compare the weak-pairing result $\epsilon\propto \phi$ to the energy spectrum $E\propto m$ of the pairing Hamiltonian (\ref{rspace}) in the weak-pairing phase. 

\begin{figure}
\includegraphics[width=3.6in]{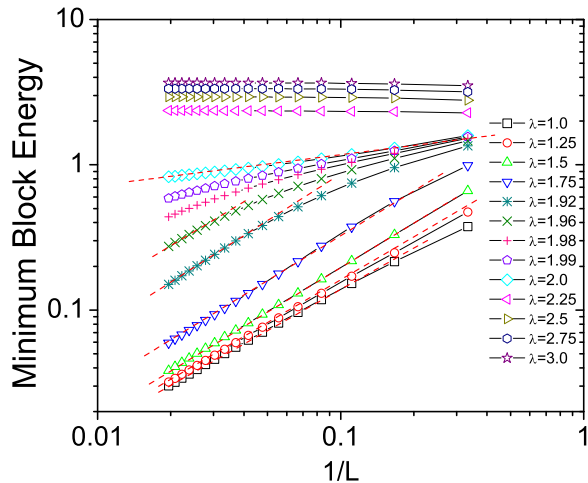}
\caption{(Color online) Finite-size scaling of the minimum quasi-particle block energy $\epsilon_1$ plotted on a log-log scale at fixed pairing amplitude $\gamma=1.0$.  In the weak-pairing phase $\lambda<2.0$, the dashed lines are best fits to the scaling form $\epsilon_1\sim1/L$.}

\label{spectrum}
\end{figure}

To test the identification further, we show in Fig.~\ref{spectrum}, the finite-size scaling of the minimum quasi-particle block energy $\epsilon_1$ plotted on a log-log scale at fixed pairing amplitude $\gamma=1.0$.  For the pairing Hamiltonian (2) on a disc of radius $R$, the minimum quasi-particle energy scales as $E_1\propto 1/R$ in the weak-pairing phase and tends to a constant in the strong-pairing phase, as $R\rightarrow\infty$.  In Fig.~\ref{spectrum}, the data for the strong-pairing phase $\lambda>\lambda_c$ tend to a constant as $L\rightarrow\infty$.  By contrast, in the weak-pairing phase $\lambda<2.0$, the minimum block energy drops to zero $\epsilon_1\sim 1/L$, for system sizes $L\gg \xi$ large compared to the diverging length scale $\xi=2\gamma/|\lambda-\lambda_c|$ characterizing critical fluctuations.  In the quantum critical regime, $L\ll \xi$, the finite-size scaling of the minimum quasi-particle energy is intermediate between those of weak and strong-pairing phases.  Remarkably, the contrast in Fig. 4 between the finite-size scaling of the weak-pairing and strong-pairing phases occurs even for relatively small block sizes $L/\xi\approx 1$.  On the other hand, the data in Fig. 2(c) and (d) show that the finite-size corrections to the entanglement entropy require significantly larger systems $L/\xi\approx 3$ to see the asymptotic behavior.  


\section{Conclusion} In this Letter, we study topological order in paired states of fermions with parity and time-reversal symmetry breaking.  Large-scale numerical calculations of the entanglement spectrum and entanglement entropy reveal universal behavior.  In particular, we find a chiral, gapless Majorana fermion excitation in the entanglement spectrum of the weak-pairing phase, and contrast this with the gapped spectrum in the strong-pairing phase.  A variety of topological phases can be described by a pairing Hamiltonian that neglects order parameter fluctuations.  We suggest that large-scale numerical calculations of the entanglement spectrum are a robust way to detect and characterize non-Abelian topological order in the ground-state wavefunction of such phases.


NBA acknowledges the 2008 Boulder Summer School and NCTS for their hospitality during the completion of this work, and support by NSF (DMR-0703992).  Computational facilities have been generously provided by HPCC at USC.  We are grateful for fruitful discussions with A. Feguin, M.P.A. Fisher, A. Kitaev, F.D.M Haldane, Z. Nussinov, K. Raman, and P. Zanardi.


 \end{document}